\documentclass[seceqn,aoas,MSNbibl,nameyear,dvips]{arximspdf}
\usepackage{dcolumn,upgreek,url,breakurl}
\usepackage{graphicx}

\doi{10.1214/14-AOAS731} 
\volume{8}
\issue{2}
\pubyear{2014}
\firstpage{1209}
\lastpage{1231}

\makeatletter

\newcommand{\ND}{\operatorname{ND}}
\newcommand{\Exp}{\operatorname{Exp}}
\newcommand{\Beta}{\operatorname{Beta}}
\newcolumntype{d}[1]{D{.}{.}{#1}}

\makeatother

\begin{document}
\begin{frontmatter}

\title{A new method of peak detection for analysis of~comprehensive
two-dimensional gas chromatography mass spectrometry data\thanksref{TT1}}
\runtitle{Peak detection in GC$\times$GC-MS}
\thankstext{TT1}{Supported in part by NSF Grant DMS-13-12603, NIH
Grants 1RO1GM087735, R21ES021311 and P30 CA022453.}

\begin{aug}
\author[a]{\fnms{Seongho}~\snm{Kim}\corref{}\thanksref{m1}\ead[label=e1]{kimse@karmanos.org}},
\author[b]{\fnms{Ming}~\snm{Ouyang}\thanksref{m2}},
\author[c]{\fnms{Jaesik}~\snm{Jeong}\thanksref{m3}},
\author[d]{\fnms{Changyu}~\snm{Shen}\thanksref{m4}}
\and
\author[e]{\fnms{Xiang}~\snm{Zhang}\thanksref{m5}\ead[label=e5]{xiang.zhang@louisville.edu}}
\runauthor{S. Kim et al.}
\affiliation{Wayne State University\thanksmark{m1}, University of Massachusetts
Boston\thanksmark{m2}, Chonnam National University\thanksmark{m3},
Indiana University\thanksmark{m4} and University
of Louisville\thanksmark{m5}}
\address[a]{S. Kim\\
Biostatistics Core\\
Karmanos Cancer Insitute\\
Wayne State University\\
Detroit, Michigan 48201\\
USA\\
\printead{e1}} 
\address[b]{M. Ouyang\\
Department of Computer Science\\
University of Massachusetts Boston\\
Boston, Massachusetts 02125\\
USA}
\address[c]{J. Jeong\\
Department of Statistics\\
Chonnam National University\\
Gwangju 500757\\
Korea}
\address[d]{C. Shen\\
Department of Biostatistics\\
Indiana University\\
Indianapolis, Indiana 46202\hspace*{41pt}\\
USA}
\address[e]{X. Zhang\\
Department of Chemistry\\
University of Louisville\\
Louisville, Kentucky 40292\\
USA\\
\printead{e5}}
\end{aug}

\received{\smonth{12} \syear{2013}}
\revised{\smonth{2} \syear{2014}}

%
\begin{abstract}
We develop a novel peak detection algorithm for the analysis of comprehensive
two-dimensional gas chromatography time-of-flight mass spectrometry
(GC$\times$GC-TOF MS)
data using normal--exponential--Bernoulli (NEB) and mixture probability
models. The algorithm
first performs baseline correction and denoising simultaneously using
the NEB model, which also
defines peak regions. Peaks are then picked using a mixture of
probability distribution to deal
with the co-eluting peaks. Peak merging is further carried out based
on the mass spectral similarities
among the peaks within the same peak group. The algorithm is evaluated
using experimental data to study
the effect of different cutoffs of the  {conditional}
Bayes factors and the effect of
different mixture models including Poisson, truncated Gaussian,
Gaussian, Gamma and exponentially modified
Gaussian (EMG) distributions, and the optimal version is introduced
using a trial-and-error approach.
We then compare the new algorithm with two existing algorithms in
terms of compound identification. Data
analysis shows that the developed algorithm can detect the peaks with
lower false discovery rates than the
existing algorithms, and a less complicated peak picking model is a
promising alternative to the more
complicated and widely used EMG mixture models. 
\end{abstract}

%

\end{frontmatter}

\section{Introduction}\label{sec1}
Multiple analytical approaches such as liquid chromatography mass
spectrometry (LC-MS), gas chromatography mass spectrometry (GC-MS) and
nuclear magnetic resonance spectroscopy (NMR) have been employed for
the comprehensive characterization of metabolites in biological
systems. One such powerful approach is comprehensive two-dimensional
gas chromatography time-of-flight mass spectrometry (GC$\times$GC-TOF
MS). Unlike other existing analytical platforms, GC$\times$GC-TOF MS
provides much increased separation capacity, chemical selectivity and
sensitivity for the analysis of metabolites present in complex samples.
The information-rich output content of GC$\times$GC-TOF MS data has
huge potential in metabolite profiling, identification, quantification
and metabolic network analysis [\citet{1};
\citet{7}; \citet{8}; \citet{15};
\citet{17}].

Metabolites analyzed on a GC$\times$GC-TOF MS system are first
separated on two-dimensional GC columns and then subjected to a mass
spectrometer, which is usually equipped with an electron ionization
(EI) ion source. The EI process fragments the metabolite's molecular
ions and results in a fragment ion mass spectrum. For metabolite
identification and quantification using the EI mass spectra, the first
step is to reduce the instrument data, that is, a collection of EI mass
spectra, to chromatographic peaks. To help readers understand the
GC$\times$GC-TOF MS data, a brief introduction is given in the
Supplementary Information [\citet{supp}].

Although numerous algorithms have been developed for peak detection in
one-dimensional GC-MS data  {[\citet{4}; \citet{12}; \citet{13}; \citet{20}]}, there are only four algorithms
available for the two-dimensional GC-MS [\citet{14};
\citet{16}; Viv\'{o}-Truyols (\citeyear{18})], including
commercial software ChromaTOF from LECO company. However, none of these
software packages is publicly available yet, except ChromaTOF that is
commercially embedded in the GC$\times$GC-TOF MS instrument. It is
therefore highly desirable to develop publicly available peak detection
methods for the analysis of GC$\times$GC-TOF MS data. On the other
hand, the existing peak detection algorithms for the analysis of
GC$\times$GC-TOF MS data perform baseline correction and denoising
separately, which may greatly increase the risk of introducing errors
from each independent stage. In fact, Wang et~al. (\citeyear{WZWL})
introduced a reversible jump Markov chain Monte Carlo (RJ-MCMC)-based
peak detection algorithm to perform simultaneous baseline, denoising
and peak identification for analysis of one-dimensional surface
enhanced laser desorption/ionization (SELDI) MS data and demonstrated
that the simultaneous process reduces false discovery rate.
{However, in practice, the use of applications of
RJ-MCMC is limited due to that prior distributions should be
appropriately assigned in order to design an effective RJ-MCMC
algorithm and make posterior distributions of parameters
computationally tractable and that constructing an MCMC chain is in
general computationally extensive.}
\citet{20} compared the performance of peak detection
among several peak detection algorithms for one-dimensional
matrix-assisted laser desorption/ionization (MALDI) MS data and showed
that the continuous wavelet-based (CWT) algorithm, which has
simultaneous baseline and denoising, provides the best performance, but
it is still a major challenge to compute accurate peak abundance
(area), which is very important in compound quantification, due to its
model-free approach. Furthermore, the existing algorithms require a
manually assigned signal-to-noise ratio (SNR) threshold and/or
denoising parameters that are usually not optimized in the existing
peak detection algorithms, resulting in a high rate of false-positive
and/or false-negative peaks.

To avoid the aforementioned difficulties in analysis of GC$\times
$GC-TOF MS data, we propose a novel peak detection algorithm using a
normal--exponen\-tial--Bernoulli (NEB) model and mixture probability
models, and the developed R package \textit{msPeak} is publicly-available
at \url{http://mrr.sourceforge.net}. The developed algorithm is
composed of the following components: (i) the proposed NEB model
performs simultaneous baseline correction and denoising, followed by
finding the potential peak regions using a
{conditional} Bayes factor-based statistical test, (ii)~the peak
picking and area calculations are carried out by fitting experimental
data with a mixture of probability model, and (iii) the detected peaks
originating from the same compound are further merged based on mass
spectral similarity. The advantages of the proposed method are that the
proposed NEB-based preprocessing requires no manually assigned SNR
threshold and denoising parameters from users, which makes it easy to
use; and, instead of searching for the potential peaks using the entire
data, the proposed algorithm reduces the entire data to peak regions
using a  {conditional} Bayes factor of the test,
eliminating the possible computational burden as well as improving the
quality of peak abundance (area). The developed algorithm is further
compared with two existing algorithms in terms of compound identification.

Besides, we investigated the performance of several probability mixture
models for peak picking based on peak regions identified by the NEB
model. It has been known that the model-based approach measures more
accurately peak abundance (area) and the exponentially modified
Gaussian (EMG) probability model performs well for fitting asymmetric
chromatographic peaks and the detection of peak position [Di Marco and Bombi (\citeyear{3});
Viv\'{o}-Truyols (\citeyear{18}); \citet{19}].
However, to our knowledge, there is no study to evaluate the EMG model
by comparing with other possible probability models in analysis of
GC$\times$GC-TOF MS data. To address this, we employed five probability
mixture models: Poisson mixture models (PMM), truncated Gaussian
mixture models (tGMM), Gaussian mixture models (GMM), Gamma mixture
models (GaMM) and exponentially modified Gaussian mixture models
(EGMM). Here PMM, GMM and EGMM were chosen based on Di Marco and Bombi's
(\citeyear{3}) work, and we proposed two new models, tGMM and GaMM, as
alternatives to GMM and EGMM, respectively.

{We hope our research can provide some insight on the
following computational/statistical challenges of the current peak
detection approaches: (i) baseline correction and denoising are
performed separately so that there is no way to communicate the
information with each step, (ii) user-defined input values are needed,
(iii) entire data are used for peak detection (compared to our proposed
approach that finds peak regions first and then detect peaks), and (iv)
there is no comparison analysis for the performance of different
chromatographic peak models.}

The rest of the paper is organized as follows. Section~\ref{sec2} presents the
proposed peak detection algorithm and introduces a trial-and-error
optimization of the developed algorithm. The real GC$\times$GC-TOF MS
data is described in Section~\ref{sec3}. In Section~\ref{sec4} we apply the developed
algorithm to real experimental GC$\times$GC-TOF MS data, followed by
the comparison with two existing algorithms in Section~\ref{sec5}. In Section~\ref{sec6}
we discuss the results and then conclude our work in Section~\ref{sec7}.

\section{Algorithms}\label{sec2}
The proposed peak detection algorithm consists of three components with
the following four steps: finding potential peak regions, simultaneous
denoising and baseline correction, peak picking and area calculation,
and peak merging (Supplementary Information Figure S1). The first two
steps are performed by hierarchical statistical models at a time, while
the peak picking and area calculation are carried out by model-based
approaches in conjunction with the first derivative test. The peak
merging then uses mass spectral similarities to recognize peaks
originated from the same metabolite. The selection of the cutoff value
of the  {conditional} Bayes factor and the probability
model is further optimized by a trial-and-error optimization.
{On the other hand, when multiple samples are analyzed, each
sample will generate a peak list after peak detection. A cross sample
peak list alignment is needed to recognize chromatographic peaks
generated by the same compound in different samples. Current existing
peak-based methods generally perform peak alignment using the mass
spectral similarity and peak position (location) distance [e.g., \citet{8}]. This peak alignment process will generate a
matched peak table for downstream data analysis, such as quantification
and network analysis. In this regard, the peak detection process plays
an important role in generating the peak list. It should be noted that,
due to either systematic (technical) or biological variations, some
peaks (molecules) may not be detected in all samples, resulting in an
incomplete peak table. There are several remedies to deal with the
issue, such as ignoring missing data, filling in zero or
imputing/estimating missing data [e.g., \citet{5};
\citet{9}].}
\subsection{Finding peak regions}\label{sec2.1}

Newton et~al. (\citeyear{11}) proposed a hierarchical approach to the
microarray data analysis using the gamma--gamma--Bernoulli model. Their\vadjust{\goodbreak}
purpose was to detect the genes that are differentially expressed. In
this study, we adopted their idea to simultaneously perform baseline
correction and denoising, by replacing the gamma--gamma--Bernoulli model
with the normal--exponential--Bernoulli (NEB) model. The proposed
hierarchical NEB model has three layers. In the first layer, the
observed data, $X_i$, are modeled through the normal distribution with
mean $\Theta_i+\mu$ and variance $\sigma^2$ for each $i$th total ion
chromatogram (TIC), where $i=1,\ldots, N$ and $N$ is the number of
TICs. Note that a TIC is a chromatogram created by summing up
intensities of all mass spectral peaks collected during a given scan
(or a given instrumental time). In other words, we assume that the
noise follows the normal distribution with mean zero and variance
$\sigma^2$. For simplicity, the homogeneous variance is assumed in this
model. Here $\Theta_i$ is the true signal of the observed signal $X_i$
and $\mu$ is either a baseline or a background. The true signal, $\Theta
_i$, is further modeled by the exponential distribution with mean $\phi
$ in the second layer. In the case that there is only noise, meaning
that no signal is present, the observed signal, $X_i$, is modeled only
with the background and noise signal. Consequently, $X_i$ follows the
normal distribution with mean $\mu$ and variance~$\sigma^2$.

\begin{figure}

\includegraphics{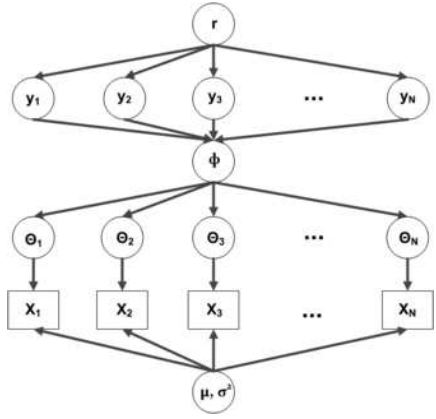}

\caption{The graphical representation of a normal--exponential--Bernoulli (NEB) model.}\label{fig:01}
\end{figure}

In this approach, we pay attention to whether the observed TIC at a
given position is significantly different from the background signal.
To do this, one more layer is introduced in the model using a Bernoulli
distribution, resulting in the NEB model. The true TICs of some
proportion $r$ are present (i.e., $\Theta_i \ne0$), while others
remain at zero ($\Theta_i = 0$). For positions where the true TIC is
present, we use the following model:
%
\begin{equation}
X_i \sim \ND\bigl(\Theta_i + \mu, \sigma^2
\bigr) \quad\mbox{and} \quad \Theta_i \sim \Exp(\phi),
\end{equation}
where $\ND$ stands for a normal distribution, $X_i$ is an observed TIC
at the $i$th position, $\Theta_i$ is the true TIC of $X_i$ of the
exponential distribution with $\phi$, and $\mu$ is the mean background
or baseline with variance $\sigma^2$. In the case that no TIC is
present, the background signal follows:
%
\begin{equation}
X_i \sim \ND\bigl(\mu, \sigma^2\bigr).
\end{equation}
Therefore, the marginal density of $X_i$ when $\Theta_i \ne0$ is
driven by
%
\begin{eqnarray}
\label{eqp1} p_1(x_i)=\frac{1}{\phi} \exp{ \biggl(
\frac{\sigma^2}{2\phi^2}- \frac
{x_i-\mu}{\phi} \biggr) \Phi \biggl(\frac{x_i-\mu-\sigma^2/\phi}{\sigma
}
\biggr)},
\end{eqnarray}
where $\Phi(\cdot)$ is the cumulative distribution function (c.d.f.) of
the standard normal distribution. When $\Theta_i=0$, the marginal
density of $x_i$ becomes the probability density function (p.d.f.) of a
normal distribution with mean $\mu$ and variance $\sigma^2$.
The detail derivation of equation (\ref{eqp1}) can be found in the
Supplementary Information. The loglikelihood $l(\mu,\sigma^2,\phi,r)$ is
%
\begin{eqnarray}
\label{loglike} &&\sum_i \bigl\{y_i \log
\bigl(p_1(x_i)\bigr)+(1-y_i)\log
\bigl(p_0(x_i)\bigr)
\nonumber
\\[-8pt]
\\[-8pt]
&&\hspace*{35pt} {}+y_i \log(r) + (1-y_i)\log(1-r)\bigr
\},
\nonumber
\end{eqnarray}
where $y_i$ is the value of the binary indicator variable, $Y_i$, at
the $i$th TIC with 0 unless there is a true significant signal and $r$
is the proportion of the true TICs. As a result, there are four
parameters ($\mu,\sigma^2,\phi,r$) along with the indicator variable
$Y_i$ ($i=1,\ldots,N$) to estimate. The graphical representation of the
proposed NEB model is depicted in Figure~\ref{fig:01}.
%

\textit{Parameter estimation}: The EM algorithm [\citet{2}] is employed to estimate the parameters of equation (\ref
{loglike}) by considering the indicator variable $Y_i$ ($i=1,\ldots,N$)
as the latent (or missing) variable. In the E-step, the latent variable
$Y_i$ ($i=1,\ldots,N$) is estimated, after fixing the parameters ($\mu,\sigma
^2,\phi,r$) at the current estimates, by
%
\begin{equation}
\hat{y}_i=P\bigl(y_i=1|x_i,\hat{\mu},\hat{
\sigma}^2,\hat{\phi},\hat{r}\bigr)=\frac
{\hat{r}p_1(x_i)}{\hat{r}p_1(x_i)+(1-\hat{r})p_0(x_i)}.
\end{equation}
To simplify the M-step, the mixture structure has been separated so
that the update estimate of $r$ is the arithmetic mean of ${\hat
{y}_i}$'s and the remaining parameters ($\mu,\sigma^2,\phi$) are
optimized by a numerical approach such as the R package \textit{nlminb}.
{It should be noted that our optimization does not
guarantee a global optimum since the R package \textit{nlminb} used is a
local optimization.} In particular, a $\Beta(2,2)$ is placed as a prior
over $r$ to stabilize the computation as well as to enable a nice
interpretation of the output, according to \citet{11},
resulting in the following $\hat r$:
%
\begin{equation}
\hat r = \frac{2+\sum_i \hat{y_i}}{2 \cdot2 + N},
\end{equation}
where $N$ is the total number of TIC. After fixing $\hat r$, the
remaining parameters are estimated by maximizing the loglikelihood as follows:
%
\begin{equation}
\bigl(\hat\mu, \hat\sigma^2, \hat\phi\bigr)=\arg\max
_{\mu,\sigma^2,\phi}l\bigl(\mu ,\sigma^2,\phi, \hat r\bigr).
\end{equation}

\textit{Finding true signals}: The true signals are found by performing
the significant test based on the posterior odds. The posterior odds of
signal at the $i$th TIC are expressed by
%
\begin{equation}
\frac{P(y_i=1|x_1,\ldots,x_N)}{P(y_i=0|x_1,\ldots,x_N)}.
\end{equation}
The posterior probability of $y_i$ given the entire TIC is
%
\begin{eqnarray}
\label{posodd} P(y_i|x_1,\ldots,x_N) & = &
\int_0^1 p(y_i,r|x_1,
\ldots,x_N)\,dr
\nonumber
\\
& = & \int_0^1 P(y_i|r,x_1,
\ldots,x_N)p(r|x_1,\ldots,x_N)\,dr
\\
& = & \int_0^1 P(y_i|r,x_i)p(r|x_1,
\ldots,x_N)\,dr
\nonumber
\end{eqnarray}
by conditional independence of the data at different TICs given the
parameter $r$. 
Using the approach in \citet{11}, the posterior odds
can be approximated by
%
\begin{equation}
\frac{p_1(x_i)}{p_0(x_i)} \frac{\hat r}{1-\hat r}, \label{odds}
\end{equation}
where $w$ can be called  {conditional} Bayes factors
since the prior odds equal unity, $P(y_i=1|x_1,\ldots,x_N)\approx
p_1(x_i)\hat r$ and $P(y_i=0|x_1,\ldots,x_N)\approx p_0(x_i)(1-\hat
r)$, by approximating equation (\ref{posodd}) at the model $r=\hat{r}$.
According to Jeffreys' (\citeyear{6}) scales, the three values of 1, 10 and 100
are used to interpret the posterior odds, meaning that the TICs are
selected using three cutoff values. That is, if the posterior odds of a
TIC are less than a selected cutoff value, then this TIC is considered
as a noise by fixing it at zero (i.e., $\Theta_i = 0$). Otherwise, the
TIC will be preserved for future analysis (i.e., $\Theta_i \ne0$).

\subsection{Denoising and baseline correction}\label{sec2.2}
Once the significant TICs (true signals) are detected by the posterior
odds, the baseline correction and denoising are performed
simultaneously based on the estimated parameters. That is, under the
assumption that the TIC is the true signal ($\Theta_i\neq0$), the
convoluted TIC, $\hat{x}_i$, is predicted by the expected true TIC
(signal) given an observed TIC, that is, $E(\Theta_i|x_i)$. By doing
the calculation described in the Supplementary Information, we can
obtain the convoluted TIC, $\hat{x}_i$, as follows:
%
\begin{eqnarray}
\label{eqdeconv} \hat x_i&=& x_i- \biggl(\hat{\mu}+
\frac{\hat{\sigma}^2}{\hat{\phi}} \biggr)
\nonumber
\\[-8pt]
\\[-8pt]
&&{} + \hat{\sigma} {\varphi \biggl(\frac{x_i-(\hat{\mu}+\hat{\sigma
}^2/\hat{\phi})}{\hat{\sigma}} \biggr)} /{\Phi \biggl(
\frac{x_i-(\hat{\mu
}+\hat{\sigma}^2/\hat{\phi})}{\hat{\sigma}} \biggr)},
\nonumber
\end{eqnarray}
where $\varphi$ and $\Phi$ are the probability density and cumulative
distribution functions of the standard normal distribution $\ND(0,1)$,
respectively.

\subsection{Peak detection and area calculation}\label{sec2.3}
After denoising and baseline correction, the vector of predicted TICs,
$\{\hat x_i\}_{i=1,\ldots,N}$, is converted into the $N_1$ by $N_2$
matrix $D$ ($=[d_{kl}]_{k=1,\ldots,N_1;l=1,\ldots,N_2}$), where $N=N_1
\cdot N_2$. Here the sizes of the rows and columns of the matrix $D$
are the same as the intervals of the first and the second dimension
retention times, respectively. In order to detect peaks for each
significant peak region, we employ five different model-based
approaches along with the first derivative test (FDT): Poisson mixture
models (PMM), truncated Gaussian mixture models (tGMM), Gaussian
mixture models (GMM), Gamma mixture models (GaMM) and exponentially
modified Gaussian mixture models (EGMM).

The peak area is calculated based on the highest probability density
(HPD) regions of 95$\%$ for model-based approaches. We assume that
$D_k$ is the $k$th row vector of the matrix $D$, where the size of
$D_k$ is $N_2$ and $1 \le k \le N_1$, and $D_k=\{d_{kl} \}_{l=1,\ldots
,N_2}$. In other words, $D_k$ is a vector of intensities at each second
dimension retention time given the $k$th first dimension retention
time. Then $D_k$ is partitioned into several nonzero peak regions, that
is, $D_k:=Z_k=\{Z_k^1,Z_k^2,\ldots,Z_k^{M_k} \}$, where $1 \le k \le
N_1; Z_k^m=\{z_{kl} \}_{l=1,\ldots,N_2^{m,k} }; 1\leq m\leq M_k;
N_2\geq\sum_{m=1}^{M_k}N_2^{m,k}$. It is noteworthy that $Z_k^m$ is a
nonzero peak region and its intensities $z_{kl}$'s are always nonzero.

\textit{First derivative test (FDT)}: FDT is used to infer the maximum
number of components (peaks) of the mixture probability models. The
first derivative is calculated over the converted nonzero vector
$Z_k^m$ with respect to the second dimension retention times to find a
peak, for the given $k$th first dimension retention time and the $m$th
nonzero peak region where $1\le k \le N_1$ and $1 \le m \le M_k$. By
doing so, the local maxima are found with respect to the second
dimension retention time. That is, for each converted vector $Z_k^m$,
we examine $z_{kl}$ whether it is a local maximum with respect to the
second dimension retention time as follows:
%
\begin{equation}
I_{kl}=\cases{ 1 &\quad if $z_{k(l-1)}-z_{kl} < 0$ and
$z_{kl}-z_{k(l+1)}>0$,
\cr
0 &\quad otherwise, } %
\end{equation}
where $I_{kl}$ is the indicator variable for peaks detected using the
second dimension retention time; 1 indicates a local maximum. In fact,
we observed that the first derivative test with respect to the first
dimension retention time has little information in most cases, due to
the relatively large value of the modulation period compared to the
chromatographic peak width in the first dimension GC. For this reason,
we used only the second dimension retention time for peak picking.
{It is noteworthy to mention again that we use the FDT
only for guessing the maximum number of the possible peaks and for an
initial value of the optimization of the model-based approach.}

\textit{Model-based approach}: The five different probability models
including Poisson, truncated Gaussian, Gaussian, Gamma and
exponentially modified Gaussian distributions are employed along with
mixture models. For each converted nonzero vector $Z_k^m =\{z_{kl} \}
_{l=1,\ldots,N_2^{m,k}}$, a mixture of the selected probability models
are fitted to the observed, convoluted intensities. Suppose $f_s (\cdot
|\xi_s)$ is a p.d.f. with a parameter of $\xi_s$ of the $s$th component,
where $1\leq s \leq S,  \xi_s$ is a scalar or a vector, and $S$ is the
number of mixture components. Then the converted nonzero intensity
$z_{kl}$ at the $k$th first dimension retention time is modeled as follows:
{
%
\begin{eqnarray}
z_{kl} \sim \ND \Biggl\{\sum_{s=1}^S
w_s \cdot f_s(t_l|\xi_s),
\tau^2 \Biggr\},
\end{eqnarray}
where $t_l$ is the second dimension retention time at the $l$th
position, $w_s$ is the non-negative weight factor of the $s$th
component, $\sum_{s=1}^S w_s = 1$, and $\ND$ stands for a normal
distribution with variance $\tau^2$. The parameters $(\zeta,S)$ are
estimated by minimizing $-$2 times loglikelihood function:
%
\begin{eqnarray}
\bigl(\hat{\zeta},\hat{S},\hat{\tau}^2\bigr)&=&\arg\min
_{\zeta,S,\tau^2} \Biggl[N_2^{m,k} \log\bigl(2\pi
\tau^2\bigr)
\nonumber
\\[-8pt]
\\[-8pt]
&&\hphantom{\arg\min_{\zeta,S,\tau^2} \Biggl[} {}+\frac{1}{\tau^2} \sum
_{l=1}^{N_2^{m,k}} \Biggl\{z_{kl}-\sum
_{s=1}^S w_s \cdot
f_s(t_l|\xi_s) \Biggr\}^2
\Biggr],
\nonumber
\end{eqnarray}
where $\hat{\zeta}=(\hat{w}_s, \hat{\xi}_s)_{s=1}^{\hat{S}}$, and $\hat
{w}_s$ and $\hat{\xi}_s$ are the estimated parameters of the $s$th
component. The p.d.f. and its parameters for Poisson, truncated Gaussian,
Gaussian, Gamma and exponentially modified Gaussian can be found in the
Supplementary Information.
}

Then 95$\%$ highest probability density (HPD) intervals are calculated
with the estimated $\hat w_s$ and $\hat\xi_s$ for each $k$th peak, $1
\le s \le\hat S$, and the length of its 95$\%$ HPD interval is
assigned as the peak area. As mentioned before, we consider the number
of peaks detected by FDT as the maximum number of peaks that can be
detected. Therefore, the number of mixture components estimated by each
model-based approach ($\hat{S}$) always becomes less than or equal to
the number of peaks detected by FDT. Furthermore, the intensities are
divided by the total sum of nonzero intensities in a given peak region
for the purpose of normalization since each model is a probability distribution.

\subsection{Peak grouping and merging}\label{sec2.4}
It is likely that multiple peaks detected can be from the same compound
due to systemic variations. To correct these multiple peaks, we use the
mass spectrum (MS) information by calculating the MS similarity among
the peaks. Since it might be computationally expensive if all the
pairwise MS similarities are calculated, we first group the peaks
according to their nonzero\vadjust{\goodbreak} peak regions and then merge the peaks having
the higher MS similarity using a user-defined cutoff value, only if
these peaks are present in the adjacent nonzero peak region(s). For
instance, assuming that two peaks $z_{kl}$ and $z_{mn}$ belong to the
nonzero peak regions $Z_k^i$ and $Z_m^j$, respectively, $z_{kl}$ and
$z_{mn}$ are considered as members of the same group if these two peak
regions are adjacent to each other. Otherwise, they are assigned to
different groups. The MS similarities among all peaks within the same
group are calculated, and the peak with the highest TIC is selected as
the representative peak in the case that multiple peaks have the MS
similarities greater than the user defined cutoff value (e.g., 0.95) by
replacing the peak area with the sum of peak areas of all merged peaks.

\subsection{Optimal selection of the cutoff value and the probability model}\label{sec2.5}
As can be observed in Section~\ref{sec3}, the optimal probability mixture model
can be different according to the detected peak regions, and so can the
cutoff value of the  {conditional} Bayes factor. For
this reason, we further consider a trial-and-error optimization of the
developed algorithm in order to select the optimal cutoff value of the
{conditional} Bayes factor and the optimal probability
mixture model. To do this, three objective functions are considered,
which are mean squared error (MSE), Akaike information criterion (AIC)
and Bayesian information criterion (BIC) that were used in Section~\ref{sec3}.
That is, given a selected objective function, we first look for the
optimal probability mixture model for each detected peak region and
then find the optimal cutoff value of the
{conditional} Bayes factor by the minimum of the sums of all the
objective functions for each cutoff value. In detail, the optimal
cutoff value $\tilde{\nu}$ is selected by the following trial-and-error
optimization:
%
\begin{eqnarray}
\label{eqopt} \tilde{\nu}=\arg\min_{\nu} \Biggl\{\sum
_{k=1}^{N_1}\sum_{m=1}^{M_k}
\tilde {J}\bigl(Z_k^m,\tilde{\beta}_m|\nu
\bigr) \Biggr\}
\nonumber
\\[-8pt]
\\[-8pt]
\eqntext{\displaystyle\mbox{with } \tilde{J}\bigl(Z_k^m
\hat {\beta}_m|\nu\bigr) = \min_{\beta_m}J
\bigl(Z_k^m|\beta_m,\nu\bigr),}
\end{eqnarray}
where $\nu\in\{1,10,100\}$ is a cutoff value, $\beta_m \in$ \{PMM,
tGMM, GMM, GaMM, EGMM\} is a probability model, and $\tilde{\beta}_m =
\arg\min_{\beta_m} J(Z_k^m |\beta_m,\nu)$. In particular, the function
$J(Z_k^m|\beta_m,\nu)$ is varied with respect to the choice of the
objective function:
%
\begin{equation}
J\bigl(Z_k^m|\beta_m,\nu\bigr)=\cases{
\displaystyle %
\frac{1}{N_2^{m,k}} \cdot \mathit{SS} \qquad \mbox{for MSE};
\cr
\displaystyle N_2^{m,k}\log\bigl(2\pi\hat{
\tau}_{\beta_m}^2\bigr)+\frac{1}{\hat{\tau}_{\beta
_m}^2} \cdot \mathit{SS} + 2 \cdot\bigl|
\hat{\zeta}^{\beta_m}\bigr|\vspace*{2pt}
\cr
\hspace*{68pt} \mbox{for AIC;}
\cr
\displaystyle N_2^{m,k}\log\bigl(2\pi\hat{
\tau}_{\beta_m}^2\bigr)+\frac{1}{\hat{\tau}_{\beta
_m}^2} \cdot \mathit{SS} +
N_2^{m,k} \cdot\bigl|\hat{\zeta}^{\beta_m}\bigr| \vspace*{2pt}
\cr
\hspace*{68pt} \mbox{for BIC,} } %
\end{equation}
where $\mathit{SS}=\sum_{l=1}^{N_2^{m,k}} \{z_{kl} - \sum_{s=1}^S w_s\cdot
f_s^{\beta_m} (t_l|\hat{\xi}_s^{\beta_m})\}^2$; ($\hat{\tau}_{\beta
_m}^2$, $\hat{\zeta}^{\beta_m}$) are the parameter estimates of a given
probability model $f_s^{\beta_m}$, and $|\hat{\zeta}^{\beta_m}|$ is the
number of the parameters of a given probability model. For the case
that a certain probability model is preferred, we can just optimize the
cutoff value of the  {conditional} Bayes factor by
fixing $\hat{\beta}_m$ at a user-defined model. In addition, we can
also just optimize the probability model by fixing $\tilde{\nu}$ at a
certain cutoff value. Note that the last two approaches are
computationally less expensive.

\section{GC$\times$GC-MS data and software}\label{sec3}
To evaluate the performance of the developed peak finding algorithm, an
experimental data set generated from a comprehensive two-dimensional
gas chromatography time-of-flight mass spectrometry (GC$\times$GC-TOF
MS) was used. The sample analyzed on GC$\times$\break GC-TOF MS is a mixture
of 76 compound standards (8270 MegaMix, Restek Corp., Bellefonte, PA)
and $C_7$-$C_{40}$ $n$-alkanes (Sigma-Aldrich Corp., St. Louis, MO).
The concentration of each compound in the mixture is 2.5 $\upmu$g/mL.
The mixture was analyzed on a LECO Pegasus 4D GC$\times$GC-TOF MS
instrument (LECO Corporation, St. Joseph, MI, USA) equipped with a
cryogenic modulator. All the statistical analyses were performed using
statistical software R 2.13.1, and the
developed R package \mbox{\textit{msPeak}} is available at \url{http://mrr.sourceforge.net}.

\section{Applications}\label{sec4}
We applied the developed algorithm to the two sets of the experimental
GC$\times$GC-TOF MS data described in the previous section. The first
data set is a small region selected from the experimental data, while
the second data set is the entire experimental data. The developed
algorithm was further compared with the two existing algorithms, the
continuous wavelet-based (CWT) algorithm and ChromaTOF, in terms of
compound identification in Section~\ref{sec5}.

\subsection{Analysis of selected data}\label{sec4.1}
For illustration purpose, we selected a small region from the entire
experimental data as shown in Supplementary Information Figure S2(b).
Upon this selected data, we performed the process of peak finding
according to the algorithm described in Section~\ref{sec2}.

Figure S2(c) displays the results of the NEB model-based significant
test as well as denoising and baseline correction for the selected
data. As described in Section~\ref{sec2}, the significant test is rendered by a
{conditional} Bayes factor of the test in equation
(\ref{odds}) with the three different odds (cutoff) values such as 1,
10 and 100. Of these, the largest cutoff value is the strictest
condition in that there are fewer significant TICs. In other words,
there are more TICs of zero in the case of the odds of 100. In this
figure, the black open circles represent the original TIC, and the red
cross ``$\times$,'' green plus ``$+$'' and blue open circles indicate the
convoluted TICs of the odds of 1, 10 and 100, respectively. Figure
S2(d) is the magnified plot of the green box in Figure S2(c). Clearly,
we can see that the odds of 1 (red cross ``$\times$'') have the most
nonzero convoluted TICs as depicted in Figure S2(d). The detected peak
regions are plotted in Figure S3 of the Supplementary Information. In
Figure S3, the total number of nonzero peak regions detected is 36, 28
and 27 when the cutoff value of odds is 1, 10 and 100, respectively. As
expected, the odds of 1 have the most nonzero peak regions.

Once the nonzero peak regions were detected, we searched for the
significant peaks at each nonzero peak region using five probability
mixture models, PMM, tGMM, GMM, GaMM and EGMM. As mentioned before, the
number of peaks detected by FDT is considered as the maximum number of
peaks to be fitted by each of five probability mixture models. Each of
PMM, tGMM, GMM, GaMM and EGMM has ($S\cdot1+1$), ($S\cdot2+1$), ($S\cdot
2+1$), ($S\cdot2+1$) and ($S\cdot3+1$) parameters to estimate, where
$S$ is the number of mixture components. Note that the truncated
Gaussian distribution has four parameters including the lower and the
upper bounds, but, in this study, these bounds are fixed with the
starting and ending indices of a given nonzero peak region. Only when
the number of data points for a given nonzero peak region is more than
or equal to the required number of parameters for a selected
probability model, then the normalized intensities are fitted using the
probability model.

To evaluate the performance of each probability model for fitting the
normalized intensities, we consider four measures: mean squared error
(MSE), $-$2 times loglikelihood (LL), Akaike information criterion (AIC)
and Bayesian information criterion (BIC). A probability model is
considered as performing better if its MSE, LL, AIC or BIC is lower. It
is noteworthy that LL is given only as a reference and will not be
directly used for comparison since each model is not nested. The
results of fitting the normalized intensities using the proposed five
probability mixture models are reported in Table~\ref{Tab:1}.

\begin{table}[t!]
\tabcolsep=0pt
\caption{Results of fitting the normalized intensities using five
probability models}\label{Tab:1}
\begin{tabular*}{\textwidth}{@{\extracolsep{\fill}}lcccccc@{}}\hline
\multicolumn{1}{c}{\textbf{Odds}}&\textbf{Measure} & \textbf{PMM} & \textbf{tGMM}
& \textbf{GMM} & \textbf{GaMM} & \textbf{EGMM} \\ \hline
\hphantom{00}1   & MSE\protect\tabnoteref[1]{tab1} & 41.34   & 30.35   & 28.29   & 33.85   & 15.63   \\
&         & (15.03) & (19.65) & (15.21) & (15.92) & (9.64)  \\
& LL      & $-$285.42 & $-$324.10 & $-$322.91 & $-$320.51 & $-$371.78 \\
&         & (41.38) & (41.27) & (41.66) & (41.24) & (46.43) \\
& AIC     & $-$277.42 & $-$313.76 & $-$312.57 & $-$308.51 & $-$358.92 \\
&         & (40.89) & (40.81) & (41.21) & (40.43) & (46.01) \\
& BIC     & $-$116.04 & $-$112.24 & $-$111.05 & $-$61.99  & $-$110.92 \\
&         & (18.08) & (17.01) & (17.42) & (18.07) & (20.20) \\
\hphantom{0}10  & MSE     & 38.34   & 2.72    & 6.41    & 7.39    & 5.54    \\
&         & (10.52) & (0.64)  & (1.95)  & (2.52)  & (2.00)  \\
& LL      & $-$255.80 & $-$310.07 & $-$305.90 & $-$301.83 & $-$336.24 \\
&         & (39.72) & (39.88) & (40.63) & (39.57) & (44.09) \\
& AIC     & $-$248.98 & $-$300.51 & $-$296.34 & $-$292.05 & $-$324.98 \\
&         & (39.28) & (39.36) & (40.11) & (38.91) & (43.67) \\
& BIC     & $-$119.43 & $-$119.40 & $-$115.23 & $-$101.50 & $-$121.72 \\
&         & (18.77) & (15.84) & (16.56) & (17.26) & (17.19) \\
100 & MSE     & 53.75   & 11.94   & 19.25   & 21.05   & 18.03   \\
&         & (13.52) & (7.73)  & (9.70)  & (10.54) & (9.77)  \\
& LL      & $-$223.97 & $-$279.79 & $-$277.68 & $-$272.47 & $-$313.56 \\
&         & (36.80) & (39.20) & (39.85) & (37.51) & (44.15) \\
& AIC     & $-$218.05 & $-$271.12 & $-$269.01 & $-$263.58 & $-$302.60 \\
&         & (36.46) & (38.63) & (39.30) & (36.92) & (43.48) \\
& BIC     & $-$113.46 & $-$113.01 & $-$112.45 & $-$101.47 & $-$107.04 \\
&         & (18.06) & (16.28) & (15.65) & (14.15) & (15.20) \\
\hline
\end{tabular*}
\tabnotetext[1]{tab1}{The values in parentheses are empirical standard errors.}
\end{table}

In the case of odds of 1, the MSE, LL and AIC of EGMM are the lowest
and these of PMM are the largest. However, the lowest BIC happens with
PMM that has the smallest number of parameters to estimate. On the
other hand, when the cutoff value of odds is 10, the MSEs are
dramatically reduced up to five times lower than those of odds 1. In
this case, tGMM has the lowest MSE, and the largest MSE still occurs
when PMM is employed. However, EGMM has the lowest LL, AIC and BIC.
Interestingly, the overall MSE becomes worse than that of odds 1, when
the cutoff value increases to 100. Nevertheless, its trend is similar
to that of odds 10. Overall, in terms of MSE, a better fitting is
achieved when 10 is considered as the cutoff value of odds and when the
truncated Gaussian mixture model is used.

\begin{figure}
\includegraphics{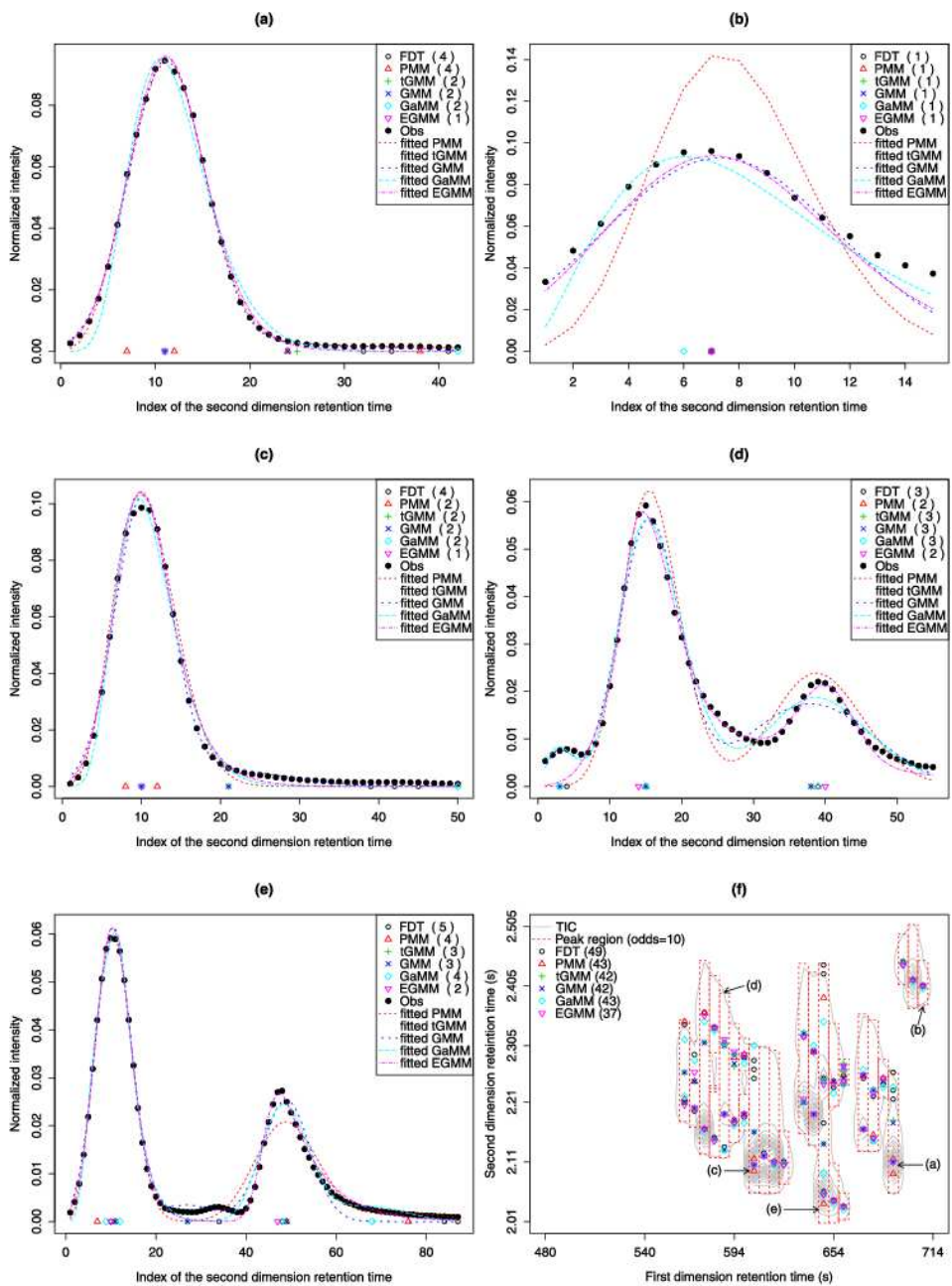}
\caption{The fitted peak regions and the detected peaks when the cutoff
value of odds is 10. 
The detected peak positions are indicated by a black circle (FDT), red
triangle (PMM), green ``$+$'' (tGMM), blue ``$\times$'' (GMM), sky-blue
rhombus (GaMM) and purple inverted-triangle (EGMM). ``Obs'' means the
observed intensities. The detected nonzero peak regions and peaks
before peak merging are in \emph{(f)}. The indices \emph{(a)--(e)} in \emph{(f)} indicate the
peak region corresponding to each of the fitted plots \emph{(a)--(e)}.}\label{fig:02}
\end{figure}

Figure~\ref{fig:02} shows the cases when each probability model
performs best among other mixture models in terms of MSE, when the
cutoff value is 10. That is, Figures~\ref{fig:02}(a)--\ref{fig:02}(e)
display the fitted curves when PMM, tGMM, GMM, GaMM and EGMM fit the
normalized intensities with the lowest MSE, respectively. In Figure~\ref{fig:02}(a), PMM has the largest number of peaks detected, which is
four as FDT does, while PMM has the lowest MSE
%
%
%
[MSE ($\times10^5$) $=$
0.06 (PMM), 0.09 (tGMM), 0.10 (GMM), 1.19 (GaMM), 0.21 (EGMM)].
Although there is only one peak in this figure, only EGMM detected one
peak. tGMM and GMM have the peaks in a very similar position, as
expected. GaMM has two peaks detected, with one peak positioned at the
upper bound of the second dimension retention time. This peak region
corresponds to the peak region (a) in Figure~\ref{fig:02}(f). tGMM has
the lowest MSE in Figure~\ref{fig:02}(b) [MSE ($\times10^5$)~$=$ 83.01
(PMM), 3.20 (tGMM), 6.08~(GMM), 7.40 (GaMM), 5.56 (EGMM)]. In this
case, all the detected peaks are located at the same position except
for GaMM, which is shifted to the left. PMM has much larger MSE,
indicating the worst fitted curve to the data points. Figure~\ref{fig:02}(c) has a similar peak shape to that of Figure~\ref{fig:02}(a),
while GMM is the best fitted model and is marginally better than tGMM
at the third decimal point in this case [MSE ($\times10^5$)~$=$ 0.88
(PMM), 0.36 (tGMM), 0.36 (GMM), 0.69 (GaMM), 0.54 (EGMM)]. Likewise,
all the probability models detected two or more peaks except for EGMM,
even though there is only one chromatographic peak. Its index of peak
region is (c) as shown in Figure~\ref{fig:02}(f). Three chromatographic
peaks are observed in Figure~\ref{fig:02}(d). PMM and EGMM detect no
peak in the beginning of the curve, while the other models correctly
detect this peak. GaMM has the lowest MSE [MSE ($\times10^5$)~$=$ 1.90
(PMM), 0.75 (tGMM), 0.76 (GMM), 0.46 (GaMM), 0.62 (EGMM)]. The case
that EGMM has the lowest MSE is depicted in Figure~\ref{fig:02}(e) [MSE
($\times10^5$)~$=$ 0.49 (PMM), 0.36 (tGMM), 0.35 (GMM), 0.24 (GaMM),
0.15 (EGMM)]. Only tGMM and GMM resolve the peak in the middle of the
curve, although their peak positions are shifted to the left. Overall,
EGMM fits the normalized intensities well with the smaller MSE, but
tGMM and GMM have a better capability of detecting the peaks that have
a relatively smaller peak height.

The detected peaks are indicated in the selected GC$\times$GC-TOF MS
data as shown in Supplementary Information Figure S4. In this figure,
the dotted red box represents the nonzero peak region detected from the
proposed NEB model, and the grey contour is plotted based on the
original TIC. When the cutoff value of odds is 1, the total number of
peaks detected by PMM, tGMM, GMM, GaMM and EGMM is 61, 53, 53, 60 and
47, respectively, as shown in Figure S4(a). Of these probability
models, PMM has the largest number of detected peaks and EGMM has the
smallest number of peaks detected. In the case of odds 10 and 100, PMM
and GaMM have the largest numbers of peaks detected out of five
probability mixture models, which are 43 and 40 for odds of 10 and 100,
respectively. EGMM still has the smallest number of detected peaks of
37, as can be seen in Figures S4(c) and S4(e). For illustration
purposes, the detected peaks when the cutoff is 10 are depicted in
Figure~\ref{fig:02}(f). Comparing Figure S4(a) with Figures S4(c) and
S4(e), it can be seen that the peak regions of small sizes have
vanished in Figures S4(c) and S4(e) when the odds increase. The
detected peaks after peak merging are shown in Figures S4(b), S4(d) and
S4(f). The peak merging dramatically reduced the number of detected
peaks up to one-third of the number of peaks before peak merging.

Furthermore, three interesting points can be observed. One is that the
cutoff value has little effect on the overall performance of peak
detection, although the odds of 10 give us the best average
performance. The second is that the peaks detected by each probability
model become similar to each other after peak merging. The last is that
the detected peaks by EGMM are very similar to those by PMM, after peak
merging, although PMM has the worst MSE, as can be seen in Table~\ref{Tab:1}. That is, the peak merging makes all probability models
comparable to each other in terms of the position of peaks detected.

\subsection{Analysis of entire data}\label{sec4.2}
We here applied the proposed peak detection algorithm to the entire
data. Before using the entire data, we removed the uninteresting areas
that were produced due to experimental noise. This can be seen in
Supplementary Information Figures S5(a) and S5(b). Then, we found the
nonzero TICs by the proposed NEB model as depicted in Supplementary
Information Figure S5(c). Similarly to the selected data in the
previous section, we selected the nonzero peak regions and then
detected the peaks using the proposed methods with the three cutoff
values, 1, 10 and 100, of odds. In the case of the cutoff value of odds
10 (1 and 100, resp.), the numbers of detected peaks are 230
(263 and 215), 223 (249 and 211), 225 (254 and 211), 229 (265 and 213)
and 215 (230 and 201), for PMM, tGMM, GMM, GaMM and EGMM, respectively,
before peak merging, while the numbers of detected peaks after peak
merging are 97 (104 and 96), 99 (107 and 97), 99 (105 and 96), 99 (106
and 96) and 96 (103 and 92), respectively. As before, the peak merging
reduces the variation in the number of detected peaks among different methods.

We further evaluated the performance of peak finding by compound
identification using Person's correlation. To standardize the
comparison, we focused on the identification of the 76 compound
standards from the peak list generated by each method. Since we knew
that 76 compound standards should be present in the sample, we examined
the quality of detected peaks by each mixture model through counting
the number of 76 compound standards that were identified out of the
detected peaks. Table~\ref{Tab:2} shows the results of the comparison.

\begin{table}
\tabcolsep=0pt
\caption{Results of compound identification of peaks detected
before/after peak merging}\label{Tab:2}
\begin{tabular*}{\textwidth}{@{\extracolsep{\fill}}lccccccd{3.0}@{}}\hline
&& \multicolumn{3}{c}{\textbf{Before peak merging}}& \multicolumn{3}{c}{\textbf{After peak merging}} \\[-6pt]
&& \multicolumn{3}{c}{\hrulefill}& \multicolumn{3}{c}{\hrulefill} \\
\multicolumn{1}{c}{\textbf{Odds}} & \textbf{Model} & \textbf{Standard}\protect\tabnoteref[1]{tab11}
& \textbf{Unique}\protect\tabnoteref[2]{tab2} &
\textbf{Peak}\protect\tabnoteref[3]{tab3} & \textbf{Standard} &
\textbf{Unique} & \multicolumn{1}{c}{\textbf{Peak}}\\
\hline
\hphantom{00}1& PMM & 32 & 72 & 263 & 32 & 64 & 104\\
& tGMM & 32 & 72 & 249 & 32 & 64 & 107\\
& GMM & 32 & 72 & 254 & 31 & 63 & 105\\
& GaMM & 32 & 73 & 265 & 31 & 64 & 106\\
& EGMM & 32 & 69 & 230 & 32 & 63 & 103\\ [3pt]
\hphantom{0}10 & PMM & 32 & 68 &230 & 31 & 61 & 97\\
& tGMM & 32 & 70 &223 & 31 & 62 & 99 \\
& GMM & 32 & 70 &225 & 31 & 61 & 99 \\
& GaMM & 32 & 71 &229 & 31 & 63 & 99 \\
& EGMM & 32 & 67 &215 & 32 & 61 & 96 \\ [3pt]
100 & PMM & 31 & 66 &215 & 30 & 60 & 96 \\
& tGMM & 31 & 67 &211 & 31 & 60 & 97 \\
& GMM & 31 & 65 &211 & 30 & 58 & 96 \\
& GaMM & 31 & 68 &213 & 30 & 60 & 96 \\
& EGMM & 31 & 64 &201 & 30 & 58 & 92 \\ [3pt]
\hphantom{0}10& OPT-MSE\protect\tabnoteref[4]{tab4} & 32 & 69 & 230 & 31 & 61 & 97 \\
\hphantom{00}1& OPT-AIC & 32 & 71 &255 & 32 & 64 & 104 \\
\hphantom{00}1& OPT-BIC & 32 & 70 &240 & 32 & 64 & 104 \\ \hline
\end{tabular*}
\tabnotetext[1]{tab11}{The number of 76 compound standards present in the list of ``Unique'' peaks;}
\tabnotetext[2]{tab2}{the number of unique compound names in the list of
peaks detected by each method;}
\tabnotetext[3]{tab3}{the number of peaks detected by
each method;}
\tabnotetext[4]{tab4}{the trial-and-error optimization.}
\end{table}

As expected, the odds of one have the most peaks detected for every
model. After peak merging, the total number of detected peaks decreased
up to more than 50$\%$, while the number of unique compounds identified
and the number of 76 compound standards identified have little
variation. Interestingly, the difference among models disappears when
we consider the total number of 76 compound standards identified, which
ranges from 30 to 32 across all the models.

\begin{figure}

\includegraphics{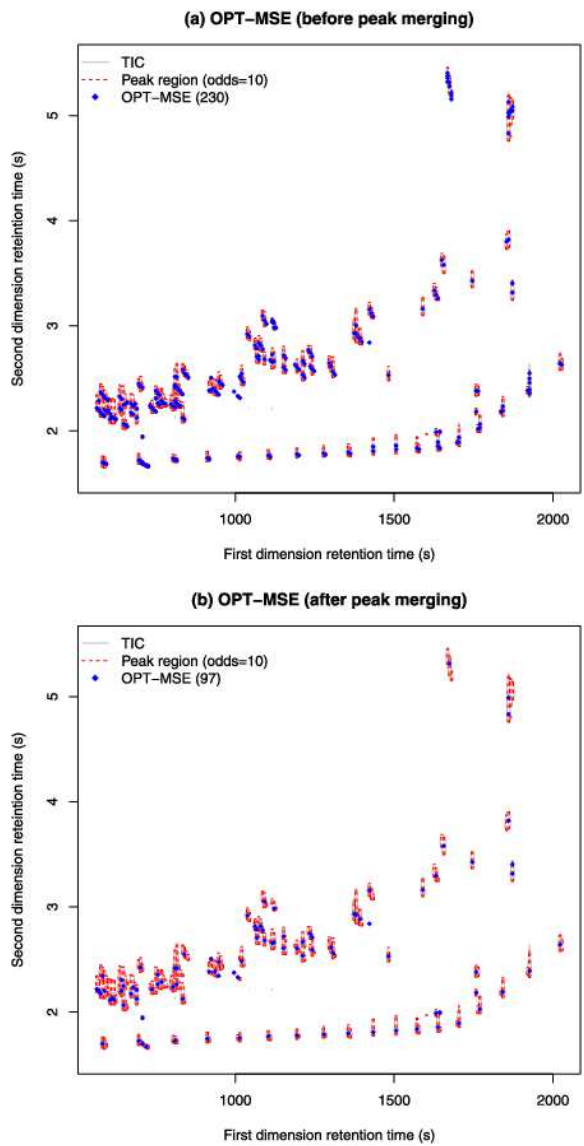}

\caption{The detected nonzero peak regions and peaks by the
trial-and-error optimization with MSE before \emph{(a)}/after \emph{(b)} peak
merging.}\label{fig:03}\vspace*{-3pt}
\end{figure}

The trial-and-error optimization was also applied to the entire data
using equation (\ref{eqopt}). The last three rows of Table~\ref{Tab:2}
show the results of the optimization with the three different objective
functions, MSE, AIC and BIC. The optimal cutoff values of odds with
MSE, AIC and BIC are 10, 1 and 1, respectively, which is consistent
with the results of the selected data (Table~\ref{Tab:1}). Since the
optimization with either AIC or BIC selects the cutoff value of one, it
detects more peaks than that with MSE. The results are shown in Figure~\ref{fig:03} when the objective function is MSE. Overall, the
optimization with MSE performs best in the sense that it finds the
smallest number of peaks but the same number of 76 compound standards.
{The peak detection results of the trial-and-error
optimization using another replicated data set can be found in
Supplementary Information Figures S7 and S8.}

%

\section{Comparison with existing algorithms}\label{sec5}
We further evaluated the developed algorithm by comparing with existing
algorithms in terms of compound identification. To do this, we employed
the two algorithms, the continuous wavelet-based (CWT) algorithm and
ChromaTOF. As mentioned earlier, there is no publicly available
software for GC$\times$GC-TOF MS data and most existing algorithms used
the one-dimensional approach for denoising and baseline correction, so
we employed CWT, which was the best-performing method based on
Yang, He and Yu's (\citeyear{20}) work. Note that ChromaTOF is the only commercial
software embedded in the GC$\times$GC-TOF MS instrument. To avoid bias
in peak merging, we used the peak list generated by each method before
peak merging. We examined the performance of CWT with the three
different signal-to-noise (SNR) ratios, 1, 2 and 3, using the
R/Bioconductor package \textit{MassSpecWavelet}. As for ChromaTOF, we used
the SNR threshold of 100. It should be noted that the unit of the SNR
threshold of ChromaTOF is different from that of CWT, so the SNR
thresholds cannot be compared directly with each other. For example,
the CWT detected no peak with SNR of 100.

Table~\ref{Tab:3} shows the results of compound identification of the entire data
using CWT and ChromaTOF before peak merging. For ease of comparison, we
added the results for the MSE-based trial-and-error optimization of
Table~\ref{Tab:2},\vadjust{\goodbreak} which performs best in Section~\ref{sec4.2}, into the table and
measured the ratios ($\%$) of the number of Standard compounds to the
number of~Unique compounds (SUR), the number of~Standard compounds to
the number of detected peaks (SPR), and the number of Unique compounds
to the number of detected peaks (UPR).

CWT detects the largest number of peaks before compound identification,
but it detects the smallest number of unique compounds as well as the
smallest number of the 76 compound standards (the true positive
compounds) identified by compound identification, resulting in the
highest false discovery rate. The commercial software ChromaTOF detects
the largest number of unique compounds and the largest number of 76
compound standards. On the other hand, although the developed algorithm
finds the smallest number of peaks, the identified number of 76
compound standards is comparable to that of ChromaTOF. Furthermore, the
highest SUR and the highest SPR are achieved by the developed
algorithm, while the highest UPR is carried out by ChromaTOF. It
suggests that the developed algorithm greatly reduces the false
discovery rate in terms of SPR. In addition, the number of detected
peaks of CWT is very sensitive to the choice of SNR threshold, while
the developed algorithm has little effect of the cutoff value of odds
on peak detection as can be seen in Table~\ref{Tab:2}. Overall, the
comparison analysis shows that ChromaTOF has much worse specificity
than our algorithm although it has better sensitivity. Moreover, CWT
has similar specificity to the developed algorithm, but the developed
algorithm has better sensitivity,

\begin{table}
\tabcolsep=0pt
\caption{Results of compound identification of peaks detected by the
developed algorithm, CWT, and ChromaTOF before peak merging} \label{Tab:3}
\begin{tabular*}{\textwidth}{@{\extracolsep{4in minus 4in}}lcd{3.0}d{4.0}d{2.2}d{2.2}d{2.2}@{}}\hline
& \multicolumn{6}{c}{\textbf{Before peak merging}} \\[-6pt]
& \multicolumn{6}{c}{\hrulefill} \\
\textbf{Method (Cutoff)} & \textbf{Standard} & \multicolumn{1}{c}{\textbf{Unique}} & \multicolumn{1}{c}{\textbf{Peak}}
& \multicolumn{1}{c}{\textbf{SUR\protect\tabnoteref[1]{tab111} ($\%$)}}
& \multicolumn{1}{c}{\textbf{SPR\protect\tabnoteref[2]{tab222} ($\%$)}}
& \multicolumn{1}{c}{\textbf{UPR\protect\tabnoteref[3]{tab333} ($\%$)}}\\ \hline
OPT-MSE (10) & 32 & 69 & 230 & 46.38 & 13.91 & 30.00 \\ [3pt]
CWT (1) & 25 & 61 & 1242 & 40.98 & 2.01 & 4.91\\
CWT (2) & 22 & 50 & 880 & 44.00 & 2.50 & 5.68\\
CWT (3) & 26 & 43 & 618 & 37.21 & 2.59 & 6.96\\ [3pt]
ChromaTOF (100) & 34 & 178 & 391 & 19.1 & 8.7 & 45.52 \\ \hline%
\end{tabular*}
\tabnotetext[1]{tab111}{Standard/Unique;}
\tabnotetext[2]{tab222}{Standard/Peak;}
\tabnotetext[3]{tab333}{Unique/Peak.}
\end{table}

\section{Discussion}\label{sec6}
To analyze GC$\times$GC-TOF MS data, we developed a new peak detection
algorithm. Unlike the existing peak detection algorithms, the proposed
algorithm performs baseline correction and denoising simultaneously
using the NEB model without any input, such as manually assigned SNR
threshold and\vadjust{\goodbreak} denoising parameters, from users. In particular, the
proposed NEB model has the ability to remove the background noise (the
series of black dots in the middle of the plot) from the raw signal as
shown in Supplementary Information Figure S5(d). Another advantage is
that it can reduce the entire data into a set of peak regions using a
statistical test of  {conditional} Bayes factors. This
is an important aspect on peak detection because the information-rich
output of MS data is usually enormous and so the processing time is one
of key bottlenecks in data analysis.

{In this work, we marginalized the MS information and
then detected the peaks using the marginalized chromatogram data (i.e.,
TIC). In our proposed peak detection, the marginalization of the MS
data is used only for the peak detection. As for the metabolite
identification, we used the non-marginalized mass spectrum for each
metabolite. In the ideal case, each peak of chromatogram (marginalized
MS data, i.e., TIC) should represent a unique metabolite (compound or
peak) in GC$\times$GC-TOF MS data. However, there are still co-eluting
metabolites from GC$\times$GC-TOF MS due to limited separation power.
For this reason, to resolve the co-eluting metabolites, ChromaTOF uses
the nonmarginalized MS data (i.e., Single Ion Chromatogram: XIC). A key
objective of using XICs of ChromaTOF is to resolve the co-eluting
metabolites, so, if there is no co-eluting peak, we think TIC should be
sufficient to detect the peaks. Furthermore, there is a much lesser
amount of metabolites co-eluting from GC$\times$GC than GC does, owing
to the increased separation power. For these reasons, we tried to
resolve the co-eluting metabolites based on TIC along with a mixture
model in the proposed approach. By doing so, although there is a
certain degree of information loss due to marginalization, we could
also see several benefits from this marginalization: (i) it reduced the
size of data, (ii) consequently, it spent less computational expense,
and (iii) it made the data much smoother [\citet{10}].
Furthermore, in comparison analysis, our approach is comparable with
ChromaTOF, which uses no marginalization of the MS data. It should be
noted that neither the commercial software ChromaTOF nor our developed
method completely detected all of the known compounds. There could be
many reasons for this, including low concentration, low ionization
frequency, inaccuracy of the peak detection algorithms, etc.}

It is a challenge to estimate the unknown number of components for a
mixture model, especially under the presence of many local optima. To
circumvent this potential issue, the developed algorithm uses the
number of peaks detected by FDT as the upper bound of the number of
components. Although this saves the computation time as well as removes
the potential difficulty, it can be a potential drawback of the
proposed algorithm when the true number of components is larger than
the number of peaks detected by FDT. However, since the FDT used in
this study was very sensitive to noise, we observed, with limited
testing data, that FDT overestimates the number of peaks in most cases
(e.g., Figure~\ref{fig:02}).

The most dominated parametric model to describe the shape of
chromatographic peaks in GC-MS and LC-MS data is an exponentially
modified Gaussian (EMG) function [\citet{3}]. In this
work, four mixture models were compared with the EMG model as can be
seen in Tables~\ref{Tab:1} and \ref{Tab:2}. The EMG model shows the
least MSE and the least number of detected peaks. However, there is no
performance difference between the mixture models in terms of
metabolite identification and detected peaks, implying that the less
complicated model has an advantage on the computation.

As mentioned, Table~\ref{Tab:2} shows no preference among different
mixture models on compound identification, although there is a clear
difference in fitting the intensities (Table~\ref{Tab:1}). For this
reason, we analyzed the MS similarity within a peak region displayed in
Figure~\ref{fig:02}(e). The results of MS similarity analysis are
depicted in Supplementary Information Figure S6. In this analysis, we
calculated the pairwise MS similarities among the data points, and then
displayed the pairwise MS similarities for each data point in Figure
S6(a). For example, a dotted line represents the MS similarities
between the $i$th point and all other $j$th points, $j\neq i$, given
the $i$th data point. Therefore, the $i$th data point must have the MS
similarity of one at the $i$th point in the plot since it is the MS
self-similarity. In this figure, we can cluster the MS similarity lines
into three groups which are the same as the number of apparent local
maxima. Furthermore, although a data point is located far from its
local maximum, its MS similarity is more than 0.8, demonstrating that
the peak region plays a more important role in compound identification
than the peak position. A similar trend can be seen in the heatmap
depicted in Figure S6(b). This can not only explain why we observe no
difference among the different models in compound identification, but
also give an insight that MS similarity-based peak detection is a
promising approach.

{Since the proposed model-based algorithms require a
fitting of the peak shape, the peak finding procedure would run a long
time. To evaluate the running time, we compared the computation time
between CWT with the cutoff of 1 and the proposed algorithm
with/without optimization with odds of 10, on a desktop with Intel Core
2 Duo CPU 3.00 GHz. The running time of CWT was 0.24 minutes, while the
proposed algorithms with PMM and optimization took 6.86 minutes and
14.10 minutes, respectively. The full optimization version of the
proposed algorithm further took 33.37 minutes. Although the proposed
algorithms relatively take more running time than CWT, it still finds
peaks in a reasonable time frame. The running time of the full
optimization approach may be a bottleneck compared to CWT, however. One
solution to speed up the computation time is to rely on parallel
computation. In general, the parallel computation is more efficient
when a lot of independent calculations are required, and the peak shape
fitting of the proposed method can be performed independently using
parallel computation for each of the peak regions.}


Even though there is not much of a difference between the developed
algorithm and the only commercial software available for GC$\times
$GC-TOF MS in terms of the number of 76 compound standards identified,
the total number of peaks detected by the proposed algorithm is about
100 peaks less than that of ChromaTOF (e.g., 230 vs. 391 in Table~\ref{Tab:3}). This is because ChromaTOF uses XIC, while the developed
algorithm uses total ion chromatogram (TIC). The XIC-based approach
requires more computation since it independently deals with each XIC of
$m/z$ values, while the TIC-based approach needs a one-time
computation. The use of XIC on the developed algorithm is left as
future work.

\section{Conclusions}\label{sec7}
We developed a novel, publicly-available algorithm to identify
chromatogram peaks using GC$\times$GC-TOF MS data, which includes three
components: simultaneous baseline correction and denoising by the NEB
model, peak picking with various choices of mixture models, and peak
merging. The proposed algorithm requires no SNR threshold and denoising
parameters from users to perform baseline correction and denoising. The
data analysis demonstrated that the NEB model-based method can detect
the peak regions in the two-dimensional chromatogram that have
chromatographic peaks, with a simultaneous baseline removal and
denoising process. Furthermore, the comparison analysis with limited
data shows that the developed algorithm can greatly reduce false
discovery rate in terms of compound identification. Among the
model-based approaches for peak picking, PMM and GMM detect more peaks,
while tGMM and EGMM have smaller MSE.  {However, there
is no apparent preference among the five model-based approaches in
terms of compound identification, and the peak shape is data-dependent.
For this reason, we further introduced a trial-and-error optimization
into the proposed algorithm to select a proper peak shape model
according to a different peak region. Among the four measures (MSE, LL,
AIC and BIC), MSE will be considered to find an optimal peak shape
model in terms of the compound identification.}


\begin{supplement}[id=suppA]
\stitle{Supplementary Information}
\slink[doi,text={10.1214/14-AOAS731SUPP}]{10.1214/14-AOAS731SUPP} 
\sdatatype{.pdf}
\sfilename{aoas731\_supp.pdf}
\sdescription{\\A~brief~introduction to GC$\times$GC-TOF MS data, derivations
of Equations (\ref{eqp1}) and (\ref{eqdeconv}), the p.d.f.s of five
probability models and Figures S1--S8 are in this Supplementary Information.}
\end{supplement}


\printaddresses

\end{document}